\documentclass[fleqn,twoside]{article}
\usepackage{espcrc2}

\usepackage{graphicx}
\usepackage[figuresright]{rotating}

\newcommand{\openone}{\leavevmode\hbox{\small1\kern-4.2pt\normalsize1}}

\newcommand{\AmS}{{\protect\the\textfont2
  A\kern-.1667em\lower.5ex\hbox{M}\kern-.125emS}}

\title{Charm and bottom baryon masses in the 1/N expansion}

\author{E. Jenkins\address[UCSD]{Department of Physics, University of California,
        San Diego, \\
        9500 Gilman Drive, La Jolla, California 92093, USA}%
        \thanks{Supported in part by the Department of Energy under grant
	DOE-FG03-97ER40546.}}
       
\begin{document}

\begin{abstract}
The masses of heavy quark baryons are studied in an expansion in $1/N_c$, $SU(3)$
flavor symmetry breaking, and heavy-quark symmetry breaking.  Very accurate
model-independent mass relations are obtained for charm and bottom baryons.
\vspace{1pc}
\end{abstract}

\maketitle

\section{INTRODUCTION}

The $1/N_c$ expansion has proven useful for studying the spin-flavor properties 
of baryons containing light quarks.  A spin-flavor symmetry for baryons is defined
in large-$N_c$ QCD.  Away from the $N_c \rightarrow
\infty$ limit, $1/N_c$ corrections which break large-$N_c$ baryon
spin-flavor symmetry can be classified in terms of the spin and flavor symmetries
which remain for finite $N_c$.  Since explicit $SU(3)$ flavor symmetry breaking 
$\epsilon \sim m_s/\Lambda_\chi$ is comparable to $1/N_c = 1 /3$ for 
QCD baryons, a combined expansion in $1/N_c$
and $\epsilon$ is necessary to explain the symmetry-breaking pattern.  The $1/N_c$
expansion has had considerable phenomenological success for $qqq$ baryons;  
$1/N_c$ suppression factors clearly are present in experimental data.

For baryons containing a single heavy quark $Q$ in HQET, there is spin-flavor
symmetry in the large-$N_c$ limit as well as in the heavy quark
limit~\cite{Jenk96}.  
The spin-flavor symmetry of heavy quark baryons with one heavy quark flavor
contains a $SU(6)_\ell$ symmetry of the light degrees of
freedom in the $N_c \rightarrow \infty$ limit, and a $SU(2)_Q$ spin symmetry
of the heavy quark in both the $m_Q \rightarrow \infty$ and 
$N_c \rightarrow \infty$ limits.
The spin-flavor symmetry of the light degrees of
freedom is broken by corrections suppressed by factors of $1/N_c$ and $SU(3)$
flavor symmetry breaking, whereas the heavy-quark spin symmetry is
broken by terms suppressed by factors of $1/N_c$ and 
heavy quark symmetry breaking $\delta_Q \sim \Lambda_{\rm QCD}/m_Q$.
Note that $SU(3)$ flavor symmetry and heavy quark spin symmetry are better
symmetries for baryons than for mesons because violation of 
spin-flavor symmetry is suppressed by additional
factors of $1/N_c$ for baryons.

Heavy quark symmetry for two heavy quark flavors $Q=c$ and $Q=b$ in HQET
generalizes to heavy-quark spin-flavor symmetry $SU(4)_Q$, 
which relates the heavy quark spin-flavor properties of charm and bottom hadrons.
Again, heavy quark spin-flavor symmetry is a better symmetry for
baryons than for mesons because heavy quark
spin-flavor symmetry violation is accompanied by
factors of $1/N_c$.

The lowest-lying spin-flavor representation for $Qqq$ baryons consists of a
completely symmetric spin-flavor representation of two light quarks combined with
a single heavy quark with $J_Q = \frac 1 2$.  Under light-quark spin and flavor, this
representation decomposes into a ${J_\ell = 0}$ flavor ${\bf {\bar 3}}$, which 
consists of the isosinglet $\Lambda_Q (Qud)$ and the isodoublet
$\Xi_c (Qsq)$ with $J= J_\ell + J_Q = \frac 1 2$, and 
a ${J_\ell = 1}$ flavor ${\bf 6}$, which consists of
$\Sigma_Q$, $\Xi_Q^\prime$, $\Omega_Q$ with $J= \frac 1 2$
and $\Sigma_Q^*$, $\Xi_Q^*$, $\Omega_Q^*$ with $J= \frac 3 2$.

The mass hierarchy of the lowest-lying charm and bottom baryon masses was predicted
in a combined expansion in $1/N_c$, $\epsilon$ and $\delta_Q$ in 
Ref.~\cite{Jenk96}.  Here,
the theoretical hierarchy 
is compared with experiment for charm baryon masses, and the predicted
pattern is seen.  Bottom baryons are predicted to obey
the same mass hierarchy with $1/m_c$ replaced by $1/m_b$.  

\section{$1/N_c$ Expansion}

In the $1/m_Q$ expansion of HQET, 
the mass of a hadron containing a single heavy quark is given by
\begin{equation}\label{mh}
M(H_Q) = m_Q + \bar \Lambda - {\lambda_1 \over {2m_Q}} - d_H {\lambda_2 \over {2
m_Q}} + \cdots  ,
\end{equation}
where  
\begin{eqnarray}
\lambda_1 &=& \langle H_Q(v) | \bar Q_v (iD)^2 Q_v | H_Q(v)\rangle , \\
d_H \lambda_2 &=& \frac 1 2 Z_Q \langle H_Q(v) | \bar Q_v g G_{\mu \nu} \sigma^{\mu
\nu} Q_v | H_Q(v) \rangle , \nonumber
\end{eqnarray}
are the matrix elements of the $1/m_Q$-suppressed operators in the heavy hadron, 
$d_H = -4 \left( J_\ell \cdot J_Q \right)$, and 
corrections of order $1/m_Q^2$ have been neglected. 
Eq.~(\ref{mh}) can be applied to mesons and to baryons, but the values of
$\bar \Lambda$, $\lambda_1$ and $\lambda_2$ will be different in the two cases.

The spin-$\frac 1 2$ ${\bf 3}$, and spin-$\frac 1 2$ and $\frac 3
2$ $\bf 6$ baryons have masses given by the $1/m_Q$ expansions   
\begin{eqnarray}
T_Q &=& {m_Q} + \bar \Lambda_T - { \lambda_{1T} \over {2m_Q}} + \cdots ,
\nonumber\\
S_Q &=& {m_Q} + \bar \Lambda_S - { \lambda_{1S} \over {2m_Q}} - 
{{4 \lambda_{2S}}
\over {2m_Q}} + \cdots , \\
S_Q^* &=& {m_Q} + \bar \Lambda_S - { \lambda_{1S} \over {2m_Q}} + 
{{2 \lambda_{2S}}
\over {2m_Q}} + \cdots, \nonumber
\end{eqnarray}
respectively.  Large-$N_c$ spin-flavor symmetry implies that the hadronic
matrix elements of these baryons are equal in the  
$N_c \rightarrow \infty$ limit.  
For finite $N_c$, the matrix elements have expansions in terms
of $1/N_c$ operators given by
\begin{eqnarray}\label{1/n}
\bar \Lambda &=& {N_c \openone} + {J_\ell^2 \over N_c}, \nonumber\\
{\lambda_1 \over {2m_Q}} &=& {{1 \over m_Q} N_Q} + 
{{1 \over N_c^2} {1 \over m_Q} N_Q
J_\ell^2}, \\
-d_H {\lambda_2 \over {2m_Q}} &=& {
{1 \over N_c} {1 \over m_Q} \left( J_\ell \cdot
J_Q \right)}, \nonumber
\end{eqnarray}
where $N_Q$ is the heavy quark number operator which is equal to $1$
for baryons containing a single heavy quark, $\openone$ is the unit operator
for baryons, and $J_\ell^i$ and $J_Q^i$ are the spins of the light degrees of 
freedom and the heavy quark, respectively. In
Eq.~(\ref{1/n}),  it is to be understood that each $1/N_c$ operator is 
accompanied by an unknown, dimensionful, ${\cal O}(1)$ coefficient, 
which has been suppressed for
simplicity.  
Eq.~(\ref{1/n}) makes a number
of interesting predictions.  For instance, $\bar \Lambda_T$ and $\bar\Lambda_S$ 
are equal at leading order $N_c$ in the $1/N_c$ expansion.  However, at
order $1/N_c$, the two matrix elements are not equal,
but are split by a contribution which is order $1/N_c^2$ relative to the leading
${\cal O}(N_c)$ term.  Similar remarks apply for the $\lambda_1$ matrix
elements.

The generalization of Eq.~(\ref{1/n}) to include $SU(3)$ flavor symmetry and 
its breaking is provided in Ref.~\cite{Jenk96}.  At the time of this work, the
$\Xi_c^\prime$ mass had not been measured, and the $1/N_c$ analysis was used to 
successfully predict $\Xi_c^\prime = 2580.8 \pm 2.1$ MeV~\cite{Jenk97}, 
to be compared with the 
subsequent experimental value $\Xi_c^\prime = 2576.5 \pm 2.3$ MeV.  
This theoretical prediction and its precision required the $1/N_c$ expansion.

Today the masses of all singly charmed baryons are measured except for the
spin-$\frac 3 2$ $\Omega_c^*$.  The most suppressed mass combination
\begin{equation}\label{cmass} 
\frac 1 4 \left[ \left( \Sigma_c^* - \Sigma_c \right) - 2 \left( \Xi_c^* -
\Xi_c^\prime \right) + \left( \Omega_c^* - \Omega_c \right) \right],
\end{equation}
which is suppressed by $\delta_Q \epsilon^2/N_c^3$ relative to the ${\cal O}(N_c)$
baryon mass, can be used to extract $\Omega_c^* =2770.7 \pm 5.9$ MeV.
Thus, it is possible to evaluate the charm baryon mass
hierarchy and compare with theory.  For bottom baryons, only
the $\Lambda_b^0$ mass is measured.  Using heavy quark spin-flavor symmetry,
it is possible to predict all of the other bottom baryon masses in 
terms of the charm baryon masses~\cite{Jenk97}.  

\begin{figure}
\includegraphics[width=15pc]{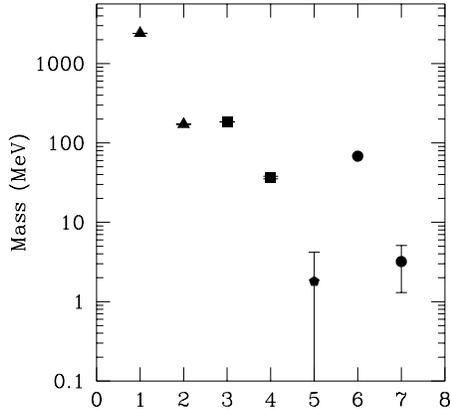}
\vspace{-18pt}
\caption{Mass hierarchy of charm baryons.}
\end{figure}

Figure 1 plots seven of the eight independent mass combinations of the 
lowest-lying charm baryon spin-flavor multiplet.  
[The mass splitting Eq.~(\ref{cmass})
is not plotted since it was used to determine the $\Omega_c^*$ mass.]  
The first mass combination
is the leading order $N_c \Lambda + m_Q$ baryon mass where $Q=c$.  The remaining mass
splittings are order $\Lambda$ times ${1 \over N_c}$, ${\epsilon}$,
${\epsilon \over N_c}$, ${\epsilon^2 \over N_c}$, ${\delta_Q \over {N_c}}$
and ${{\epsilon \delta_Q} \over {N_c^2}}$, respectively.  Fig.~1 shows that the
$1/N_c$ splitting is comparable to the $\epsilon$ splitting, and that the
$\delta_Q/N_c$ splitting is a bit larger than the $\epsilon/N_c$ splitting for
$Q=c$.
The most suppressed splittings are consistent with 
the predicted
${\epsilon^2 \over N_c}$ and ${{\epsilon \delta_Q} \over {N_c^2}}$ hierarchy
as well.

It is possible to determine the $1/m_Q$-dependent contributions to the
charm baryon mass splittings which do not violate heavy quark spin symmetry by
comparison with the analogous mass splittings for baryons containing no heavy quark.
The $1/N_c$ expansions of $qqq$ and $Qqq$
baryon masses are given in the flavor symmetry limit by
\begin{eqnarray}
M_{(qqq)} &=& N_c \openone + {1 \over N_c} J_\ell^2, \nonumber\\
M_{(Qqq)} &=& N_c \openone + {N_Q m_Q} 
+{1 \over N_c} J_\ell^2 \\
&+&{{1 \over {N_c^2 m_Q}} N_Q J_\ell^2} + 
{{1 \over {N_c m_Q}} \left( J_\ell \cdot J_Q \right)},\nonumber
\end{eqnarray}
which shows that the $N_Q$ and the $N_Q J_\ell^2$ splittings can be extracted
by making this comparision.

\begin{figure}
\includegraphics[width=15pc]{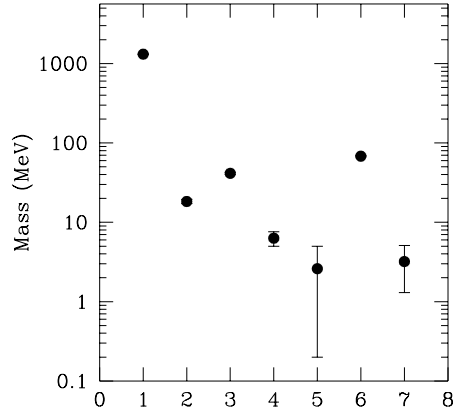}
\vspace{-18pt}
\caption{$1/m_Q$-dependent mass splittings
for charm baryons.
}
\end{figure}
Figure 2 plots
the $1/m_Q$-dependent portion of the first five mass
splittings of Fig.~1 together with the two 
heavy quark spin-violating mass splittings (points 6 and 7)
of Fig.~1.
The first mass (point 1) yields $m_Q$ or the charm quark mass at leading order in
$1/m_Q$, whereas
the six other mass splittings are order $\Lambda$ times the 
the dimensionless suppression factors
${\delta_Q \over {N_c^2}}$,
${{\epsilon\delta_Q}\over {N_c}}$,
${{\epsilon\delta_Q} \over {N_c^2}}$,
${{\epsilon^2 \delta_Q} \over {N_c^2}}$,
${\delta_Q \over {N_c}}$, and 
${{\epsilon \delta_Q} \over {N_c^2}}$, respectively.  It is interesting to
note, for example, that the
two ${{\epsilon\delta_Q} \over {N_c^2}}$ splittings (points 4 and 7) are in good 
agreement, showing that the same heavy quark symmetry-violating parameter
$\delta_Q$ is governing heavy quark spin-conserving and spin-violating
mass splittings.

\section{CONCLUSIONS}

The $1/N_c$ hierarchy of the $1/N_c$ expansion is evident in the masses of $Qqq$
baryons as well as $qqq$ baryons.  The $1/N_c$ expansion, together with $SU(3)$
flavor violation and heavy-quark symmetry violation, gives a
quantitative understanding of spin-flavor symmetry breaking for heavy quark
baryons.  An intricate pattern of spin-flavor symmetry breaking is predicted since
$1/N_c$, $\epsilon$, and $\delta_Q$ for $Q=c$ are comparable in magnitude.  The same
$1/N_c$ hierarchy is expected to appear for bottom baryon masses, and it is
possible to predict the bottom baryon mass splittings in terms of charm baryon
mass splittings.  Heavy quark spin-flavor symmetry is a better symmetry
for $Qqq$ baryons than for heavy quark mesons because violation of the spin-flavor
symmetry is suppressed by factors of $1/N_c$ as well as $1/m_Q$.

\end{document}